\documentclass{iopart}

\usepackage{iopams}
\usepackage[dvips]{graphicx}

\begin{document}


\title[A model-independent analysis...]{A model-independent 
analysis of the dependence of the
anomalous $J/\psi$ suppression on the number
of participant nucleons}

\author{A.P.~Kostyuk$^{1,2}$ and
W.~Greiner$^{1}$}
\address{$^1$ Institut f\"ur Theoretische Physik,
Universit\"at  Frankfurt, Germany}
\address{$^2$ Bogolyubov Institute for Theoretical Physics,
Kyiv, Ukraine}

\begin{abstract}
A recently published experimental dependence of the $J/\psi$ 
to Drell-Yan ratio
on the measured, by a zero degree calorimeter,
forward energy $E_{ZDC}$ in Pb+Pb collisions at the CERN SPS is
analyzed. Using a model-independent approach, it is shown  that the
data are at variance with a earlier published experimental
dependence of the same quantity on the transverse energy of
neutral hadrons $E_T$. The discrepancy is related to a moderate
centrality region: $100 \lesssim N_p \lesssim 200$ ($N_p$ is the
number of participant nucleons) and is peculiar only to the data
obtained within the `minimum bias' analysis (using the `theoretical
Drell-Yan'). This could result from systematic experimental errors
in the minimum bias sample. A possible source of the errors is
discussed.
\end{abstract}



Recently, the NA50 collaboration published new data on the
centrality dependence of the $J/\psi$ suppression pattern in Pb+Pb
collisions at CERN SPS \cite{NA50}. The centrality of the
collisions was estimated by measuring the energy of projectile
spectator nucleons $E_{ZDC}$ by a zero degree calorimeter. It was,
however, mentioned \cite{comm} that the new data may be at
variance with the ones published by the 
same
collaboration in
Refs. \cite{anomalous,threshold,evidence}, where the transverse
energy of produced neutral hadrons $E_T$ was used as a centrality
estimator. The two sets of data (for brevity, we shall call them
respectively `$E_{ZDC}$ data' and `$E_T$ data'), having very similar
qualitative behavior, seem, nevertheless, to be at variance
quantitatively. It was demonstrated in Ref. \cite{comm} that using
three completely different $J/\psi$ production models if a
model is fitted to the $E_T$ data, it does not agree with the
$E_{ZDC}$-that and vice versa. Earlier it was mentioned
\cite{capella} that the co-mover model, which describes very well
the $E_T$ data, does not show similar agreement with the
$E_{ZDC}$ set.

Although the previous analysis pointed out a possible problem, its
results can still be considered as inconclusive. Indeed, one may
believe that the problem is related to the used models rather than
to the data, and the `truly correct model' should agree with the
both data sets simultaneously.

In this letter, we use a model-independent analysis of the
data to show that {\it any} model would be at variance at least with
one of the data sets. This points out presence of systematic errors.


There are two types of data in each set, depending on the
analysis procedure used to obtain the ratio $R$ of $J/\psi$
multiplicity to the multiplicity of Drell-Yan pairs. In the {\it
standard} analysis, the number of both $J/\psi$ and Drell-Yan
events are extracted from the measured dimuon spectra, i.e. the
{\it physical} Drell-Yan is used as a reference. In the {\it
`minimum bias'} analysis, only the number of $J/\psi$ events and the
total number of collisions (minimum bias events) in each
centrality bin are found from the experiment. The ratio $R$ is
obtained by dividing the measured ratio of the number of $J/\psi$
events to the number of minimum bias events by the {\it
theoretical} ratio of Drell-Yan multiplicity to the probability of
a minimum bias collision at the corresponding centrality. Then the
overall factor is fixed using the standard analysis data. The
theoretical Drell-Yan to minimum bias ratio is given by
\begin{equation}\label{TheoDY}
\langle DY \rangle_{E_x} = \frac{ \int d^2 b P(E_x|b) \langle
DY(b) \rangle P_{int}(b) }{\int d^2 b P(E_x|b) P_{int}(b)},
\end{equation}
where $\langle DY(b) \rangle$ is the average number of Drell-Yan
dimuon pairs per Pb+Pb collision at impact parameter $b$ and 
$P_{int}(b)$ is the probability that an interaction between two
nuclei at impact parameter $b$ takes place. Both quantities are
calculated in the Glauber approach. The subscript $x$ stands for
$T$ or $ZDC$.

The $E_T$-distribution of events at fixed impact parameter $b$,
$P(E_T|b)$, is assumed to have a Gaussian form with the central
value and dispersion given by
\begin{equation}
\langle E_T(b) \rangle = q N_p(b) \ \ \ \ \ \
\sigma_{E_T}(b) = q \sqrt{a N_p(b)}.
\end{equation}
Here $N_p(b)$ is the average number of participant nucleons in a
Pb+Pb collision at impact parameter $b$ calculated in the Glauber
approach. The parameter values $q=0.274$ GeV and $a=1.27$
\cite{Chaurand} are fixed from the minimum bias transverse energy
distribution \cite{evidence}.

The $E_{ZDC}$-distribution $P(E_{ZDC}|b)$ is assumed to be a
Gaussian with the central value and dispersion given by Eqs. (1)
and (2) of Ref. \cite{NA50}.


At a given dependence $\langle J/\psi (b) \rangle$ of the average 
number of $J/\psi$'s per Pb+Pb collision on the impact 
parameter $b$,
the ratio R in a finite-size $E_x$-bin, $E_x^{min} \leq E_x \leq
E_x^{max}$ can be calculated as:
\begin{equation}\label{Rint}
R(E_x^{min} \leq E_x \leq E_x^{max}) = \frac{
\int_{E_x^{min}}^{E_x^{max}} d E_x \int d^2 b P(E_x|b) \langle
J/\psi (b) \rangle P_{int}(b)} {\int_{E_x^{min}}^{E_x^{max}} d E_x
\int d^2 b P(E_x|b) \langle DY(b) \rangle P_{int}(b) }
\end{equation}
In absence of essential systematic errors, the $E_T$- and
$E_{ZDC}$ data should be consistent with each other: the formula
(\ref{Rint}) should fit the both data sets simultaneously: the
$\chi^2$ per degree of freedom, $\chi^2/ndf$, that includes $E_T$
as well as $E_{ZDC}$ data points should be in the vicinity of $1$.
In the following we estimate in a model-independent way the lower
bound for $\chi^2/ndf$ and show that it is in fact essentially
larger than $1$.

Instead of $\chi^2$ per degree of freedom, we shall calculate the
$\chi^2$ per experimental point, $\chi^2/nep$, which provides a
lower estimate for $\chi^2/ndf$:
\begin{equation}
\chi^2/ndf \ge \chi^2/nep,
\end{equation}
because $ndf \le nep$.

We approximate the unknown function $\langle J/\psi (b) \rangle$
in the following way:
\begin{itemize}
\item We fix $J$ nodes $b_j$ on the $b$-axes: $b_0=0$, while $b_J$
is taken sufficiently large ($b_J=22$ fm) so that the probability
of interaction of two Pb-nuclei and nuclear modification of
$J/\psi$ production at $b=b_J$ are negligibly small. 
In our calculations, we fix the rest of nodes in two different ways:\\
(a) equidistant
\begin{equation}\label{deltab}
b_{j+1}-b_{j} = \Delta b = const, \ \ \ \ j=0,\dots,J-1
\end{equation}
and\\ (b) so that the difference of the average number of
participant nucleons between every two neighboring nodes is the
same:
\begin{equation}\label{deltanp}
N_p(b_{j})- N_p(b_{j+1}) = \Delta N_p = const, \ \ \ \ j=0,\dots,J-1.
\end{equation}
\item
The values of $\langle J/\psi (b) \rangle$ in all nodes except the
last one:
\begin{equation}
\langle J/\psi (b_j) \rangle \equiv x_j, \ \ \ \ j=0,\dots,J-1,
\end{equation}
are treated as free parameters. The value in the node $b_J$ is
fixed assuming the same $J/\psi$ to Drell-Yan ratio in extremely
peripheral Pb+Pb collisions as in p+p:
\begin{equation}
\langle J/\psi (b_J) \rangle \equiv x_J = 53.5\  \langle DY(b_J)
\rangle
\end{equation}
\item The value of $\langle J/\psi (b) \rangle$ between the nodes
is given by a linear interpolation:
\begin{equation}\label{lin}
\langle J/\psi (b) \rangle =
\frac{(b_{j+1}-b)x_j+(b-b_j)x_{j+1}}{b_{j+1}-b_j}, \ \ \ b_j < b <
b_{j+1}.
\end{equation}
\item There is no reason to expect oscillating dependence of the
$J/\psi$ multiplicity on the centrality. Therefore, $x_j$ are
subjected to the constraint
\begin{equation}\label{constraint}
\frac{x_{j-1}}{\langle DY(b_{j-1}) \rangle} \leq
\frac{x_{j}}{\langle DY(b_{j}) \rangle},
\end{equation}
which ensures monotonicity of the $J/\psi$ to Drell-Yan ratio $R$
as function of $b$.
\item One more constraint is needed to avoid negative values of 
$J/\psi$ multiplicity:
\begin{equation}\label{constraint0}
x_0 \geq 0.
\end{equation}
\end{itemize}
Any non-negative
function $\langle J/\psi (b) \rangle$ related to a monotonic
$R(b)$, including even a disconnected one, can be arbitrarily well
approximated in the above way, provided that $J$ is sufficiently
large. Therefore, minimizing of $\chi^2$ with respect to all
$x_j$, $0 \leq j < J$, at sufficiently large $J$ one gets the
value, which is, at least, not larger than the minimum $\chi^2$
in {\it any} reasonable
model.

Using Eqs.(\ref{Rint}) and (\ref{lin}) one gets a linear
dependence of $R$ on $x_j$:
\begin{equation}
R^{(n)}= \sum_{j=0}^{J} A_j^{(n)} x_j.
\end{equation}
Here
\begin{eqnarray}
A_0^{(n)} &=& \frac{ \int_{E_x^{(n),min}}^{E_x^{(n),max}} d E_x
a_{0}^{+}(E_x) } {\int_{E_x^{(n),min}}^{E_x^{(n),max}} d E_x
\int_0^{\infty} db  b P(E_x|b) \langle DY(b) \rangle P_{int}(b) };
\label{A0} \\
A_j^{(n)} &=& \frac{
\int_{E_x^{(n),min}}^{E_x^{(n),max}} d E_x
\left[a_{j}^{-}(E_x) + a_{j}^{+}(E_x) \right]}
{\int_{E_x^{(n),min}}^{E_x^{(n),max}} d E_x \int_0^{\infty} db b
P(E_x|b) \langle DY(b) \rangle P_{int}(b) }, \ \ j=1,\dots,J-1; \label{Aj}\\
A_J^{(n)} &=& \frac{
\int_{E_x^{(n),min}}^{E_x^{(n),max}} d E_x a_{j}^{-}(E_x)}
{\int_{E_x^{(n),min}}^{E_x^{(n),max}} d E_x \int_0^{\infty} db  b
P(E_x|b) \langle DY(b) \rangle P_{int}(b) }; \label{AJ}
\end{eqnarray}
where
\begin{eqnarray}
a_{j}^{+}(E_x) &=& \int_{b_{j}}^{b_{j+1}} d b b P(E_x|b)
\frac{b_{j+1} - b}{b_{j+1} - b_j} P_{int}(b), \ \ \ \ \
j=0,\dots,J-1;\\
a_{j}^{-}(E_x) &=& \int_{b_{j-1}}^{b_j} d b b P(E_x|b) \frac{b - b_{j-1}}{b_j -
b_{j-1}} P_{int}(b), \ \ \ \ \
j=1,\dots,J.
\end{eqnarray}

In our fitting procedure, the initial values of $x_j$  are chosen
randomly (using a random number generator). Then the $\chi^2$ is
minimized in such a way that the constraint (\ref{constraint}) is
fulfilled at each iteration step.

\begin{figure}[p]
\begin{center}
\vspace*{1cm}
\includegraphics[height=14.5cm]{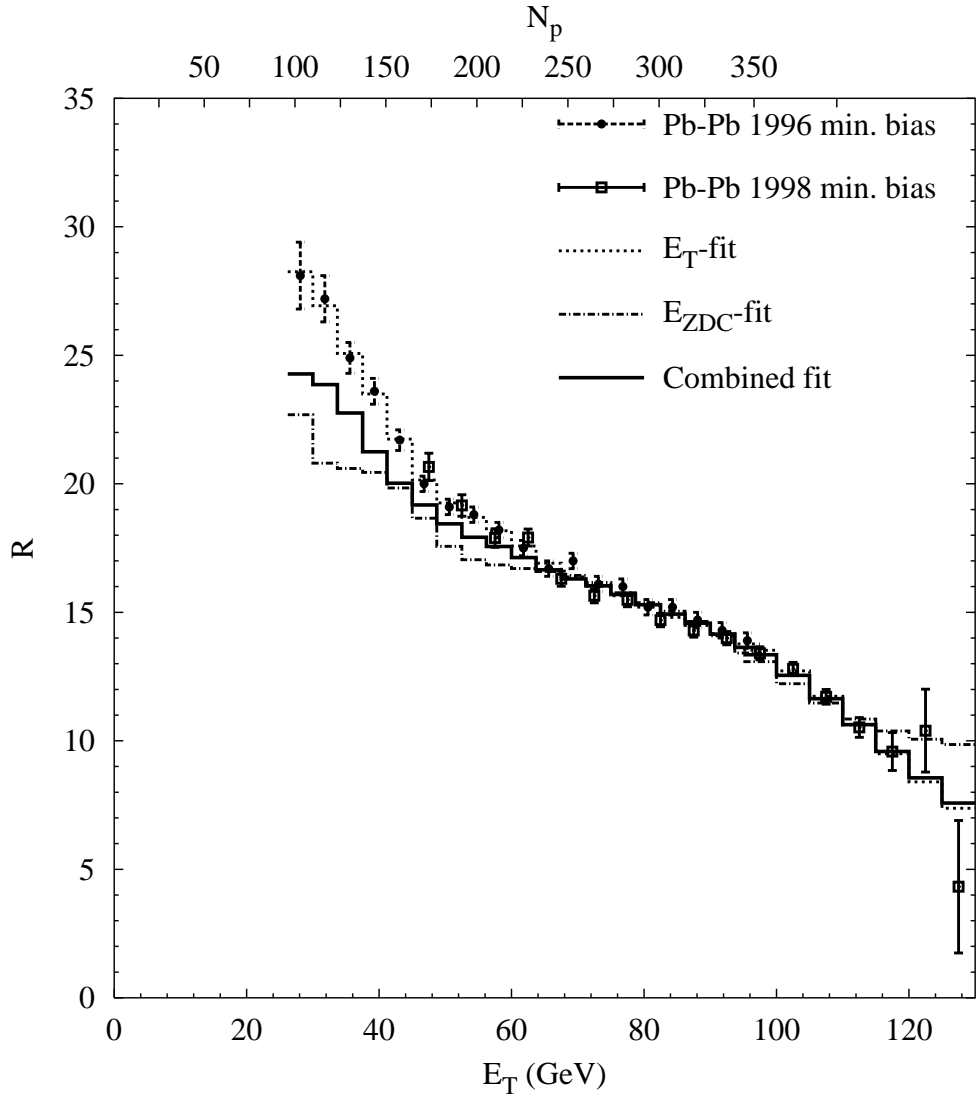}
\caption{The dependence of the $J/\psi$ to Drell-Yan ratio $R$ on
the transverse energy $E_T$. The corresponding average number of
participant nucleons $N_p$ are shown on the upper axis. The points
with error bars are the NA50 `minimum bias' analysis data (the
$E_T$ data). The dotted and dot-dashed line are fitted to the
$E_T$ data and the $E_{ZDC}$ data (see Fig. \ref{figE_ZDC}),
respectively. The solid line represents the best possible fit to
the both data sets simultaneously.}
 \label{figE_T}
\end{center}
\end{figure}

\begin{figure}[p]
\begin{center}
\vspace*{1cm}
\includegraphics[height=14.5cm]{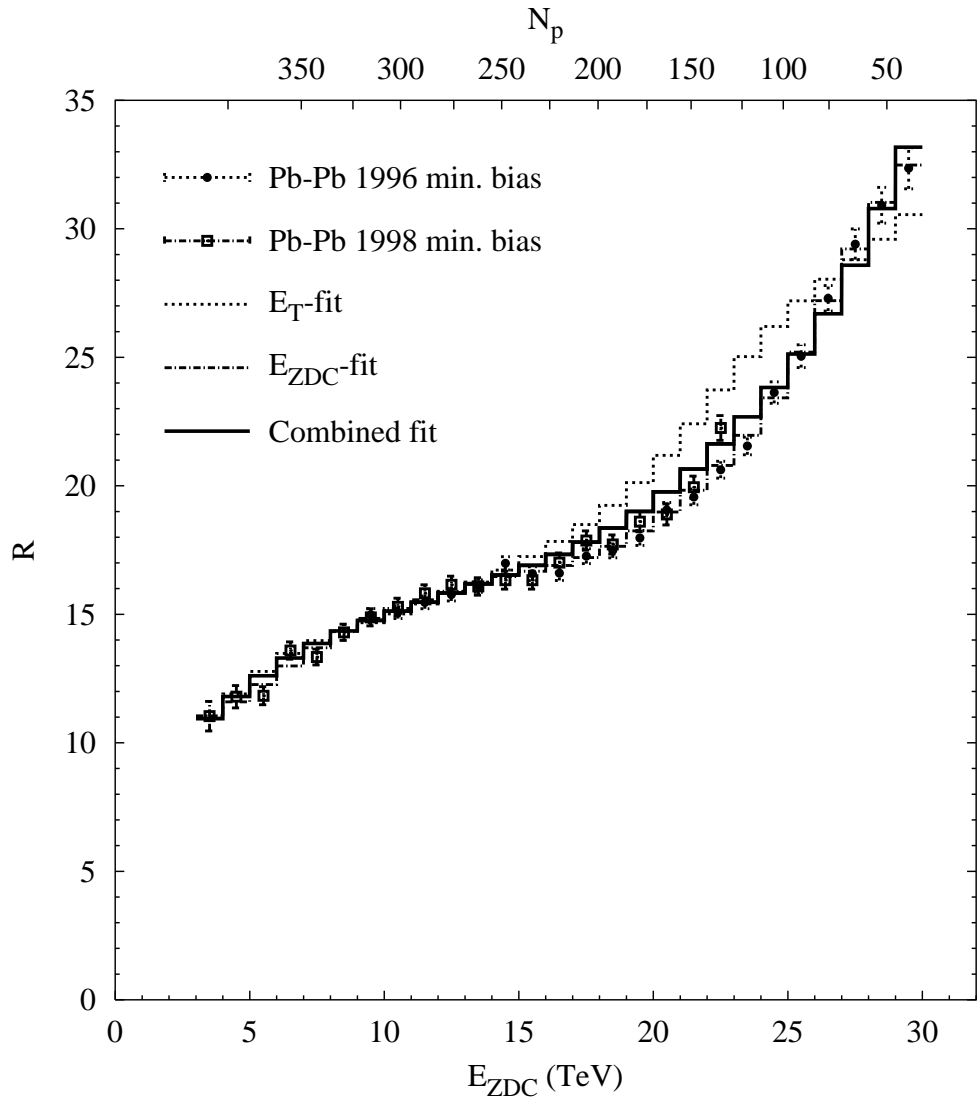}
\caption{The dependence of the $J/\psi$ to Drell-Yan
ratio $R$ on the energy of zero degree calorimeter $E_{ZDC}$.
The corresponding average
number of participant nucleons $N_p$ are shown on the upper axis.
The points with error bars are the NA50 `minimum bias' analysis data
(the $E_{ZDC}$ data). The meaning of lines is the same as in
Fig. \ref{figE_T}.}
 \label{figE_ZDC}
\end{center}
\end{figure}

The result of the fit of the NA50 {\it minimum bias} data is
presented in Tab. \ref{tab1} and in Figs.\ref{figE_T} and
\ref{figE_ZDC}. As is seen from the table at $J \ge 100$
the value of
$\chi^2/nep$ saturates and ceases to depend on the way, how 
the nodes are fixed. Therefore $J=200$ can be
accepted as sufficiently large to provide a reliable lower
estimate for $\chi^2/nep$. First, we fitted the $E_T$- and
$E_{ZDC}$ data separately (`$E_T$ fit' and the `$E_{ZDC}$ fit',
respectively). In both cases the lower estimate of $\chi^2/nep$ is
smaller than 1, which indicates that the difference
between the data from 1996 and 1998 runs within each minimum 
bias dataset can be well explained by statistical fluctuations only. 
But if one
compares the $E_T$ data and $E_T$ dependence of the $J/\psi$ to
Drell-Yan ratio, calculated with $x_j$'s found from the
$E_{ZDC}$ fit (see Fig. \ref{figE_T}), one can observe clear
discrepancy at moderate centrality ($N_p=100$--$200$). Similar
discrepancy is observed, when the $E_{ZDC}$ dependence resulting
from the $E_T$ fit is compared with the $E_{ZDC}$ data.

Finally, we tried to fit the both sets simultaneously (the
`combined fit'). The calculations show that $\chi^2/nep=3.28$.
This means that $\chi^2/ndf \geq 3.28$ in any model. Again, it is
clearly seen from Figs.\ref{figE_T} and \ref{figE_ZDC} that the
problem is related to the centrality region $N_p=100$--$200$. The
discrepancy cannot be explained by statistical fluctuations (this
hypothesis is excluded at the level of about $10^{-19}$ !).
Therefore, at least one of the data sets contains essential
systematic errors.

There are reasons to suspect that these are the $E_{ZDC}$ data
that contain the systematic errors and the source of these errors
is a distortion of the minimum bias sample. Indeed, a crucial
point of the minimum bias analysis is the assumption that both
the measured Drell-Yan multiplicity as well as the experimental 
minimum bias centrality distribution are exactly proportional to the 
corresponding theoretical values, i.e. to the numerator and 
denominator of Eq.(\ref{TheoDY}), respectively. 
The experimental and theoretical minimum bias
probabilities should cancel (up to a constant factor), 
when one divides the {\it experimental}
$J/\psi$ to minimum bias ratio by the {\it theoretical} Drell-Yan
to minimum bias ratio (\ref{TheoDY}), leaving finally only
$J/\psi$ to Drell-Yan ratio.


While the experimental minimum bias $E_T$ distribution perfectly
agrees with the Glauber model (see Fig. 1  of
Ref.\cite{evidence}),\footnote{ Disagreement is observed only in
the low $E_T$ region, because the efficiency of the target
identification algorithm is less than unity there.} 
the agreement of the $E_{ZDC}$ minimum bias distribution (see
Fig.1 of Ref.\cite{NA50}) seems to be not so perfect.
Unfortunately, the data plotted in Ref.\cite{NA50} have never been
published in numerical form and the value of $\chi^2/\mbox{dof}$
characterizing the quality of the fit of the minimum bias
$E_{ZDC}$ distribution has not been quoted. Nevertheless, it is seen
from the plot that the experimental points lie slightly above the
theoretical curves in the region $25 \gtrsim E_{ZDC} \gtrsim 18$
TeV (which corresponds to $100 \lesssim N_p \lesssim 200$).
Although the difference does not exceed even the size of the
point symbols, it, nevertheless, indicates a sizable discrepancy,
due to logarithmic scale of the vertical axis of the plot. Because
of this discrepancy, the measured and calculated minimum bias
distribution do not cancel completely, which may lead to the
observed systematic errors.

The question about the origin of the distortion of the minimum
bias distribution is not simple. The final answer can hopefully be
given after a detailed analysis of the raw experimental data
supplemented by Monte-Carlo simulations. This is out of the scope
of our present analysis. We restrict ourselves only to formulating
a `working hypothesis', which is to be tested in course of more
detailed investigations. The observed discrepancy may come from
interactions of reaction fragments with nuclei inside the zero
degree calorimeter. If such an interaction takes place, a fraction
of the produced particles (those having large transverse momenta)
may leave the calorimeter without deposing their energy in it.
This may cause losses of the zero-degree energy, which can result in
distortion of the shape of the minimum bias distribution. 
If our hypothesis is correct, the data analysis procedure should 
be modified to take into account the losses of the zero degree 
energy. Otherwise the mentioned problem will persist
even in a minimum bias analysis of  the new, year 2000 data.
 


\begin{figure}[p]
\begin{center}
\vspace*{1cm}
\includegraphics[height=14.5cm]{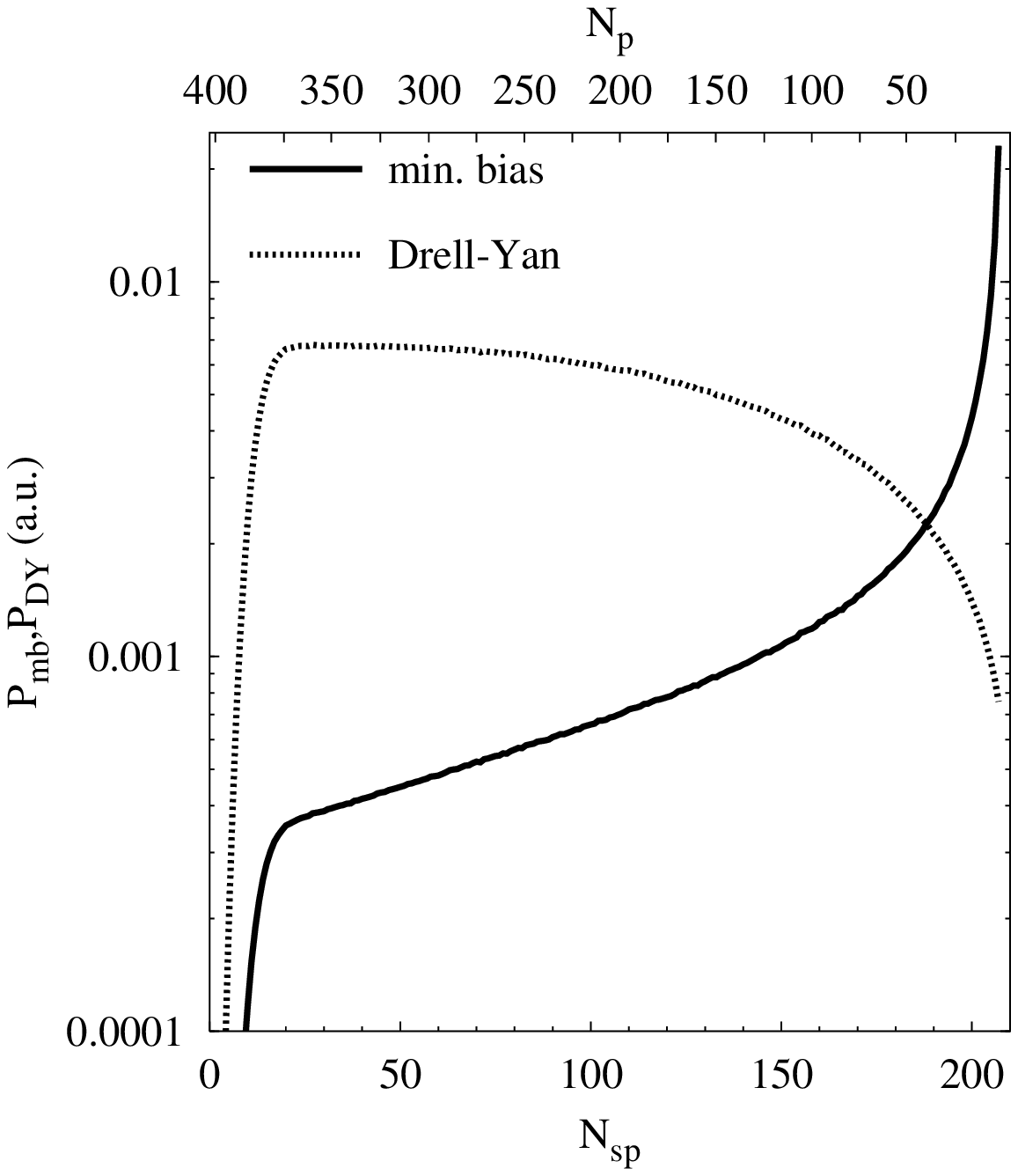}
\caption{The probability distribution of minimum bias and
Drell-Yan events as a function of the number of spectator
nucleons $N_{sp}$ in arbitrary units (a.u.), obtained within
the Glauber approach.}
 \label{distribution}
\end{center}
\end{figure}

From Fig. \ref{distribution}, one can easily see that the {\it
minimum bias} sample is much more sensitive to this effect than
the {\it dimuon} one. Indeed, if spectators from a {\it peripheral}
collision loose an essential part of their energy, this event is
registered by the apparatus as 
a {\it moderate centrality}
collision. The number of peripheral minimum bias events is about an
order of magnitude larger than the number of events at moderate
centrality. Therefore, even if the fraction of peripheral events
that suffer from losses of the zero degree energy is small, the
contamination at moderate centrality can be sizable. In contrast,
the number of Drell-Yan events at moderate centrality is larger
than the number of peripheral ones. Therefore the Drell-Yan sample
is much less influenced. This is supported by our comparison
of the standard analysis data. Even the newest data
\cite{Ramello}, which have smaller statistical errors than older
ones, do not show any contradiction between $E_T$- and the
$E_{ZDC}$ data sets: according to our analysis, the lower bound of
$\chi^2/ndf$ for the combined fit of the new standard analysis data is
$0.62$.

The region of low $E_{ZDC}$ ($E_{ZDC} \lesssim 9$ TeV), i.e. 
the region of the `second drop' of the $E_{ZDC}$ data, deserves
special attention. All models that have been successful describing
the $E_T$ data at $E_T \gtrsim 100$ GeV
\cite{capellaD,Mai,Ko:02,Grandchamp:02} explain the `second drop'
of the $J/\psi$ to Drell-Yan ratio by $E_T$-losses in the dimuon event
sample with respect to the minimum bias one and/or by the
influence of $E_T$-fluctuations on the $J/\psi$ production or
suppression. Both effects do not influence the $E_{ZDC}$ data,
this means that {\it all} the models
\cite{capellaD,Mai,Ko:02,Grandchamp:02} fail to describe the
$E_{ZDC}$ data at $E_{ZDC} < 9$ TeV. If the drop at low $E_{ZDC}$
indeed exists, the correct model should explain the
high-centrality behavior of both $E_T$- and $E_{ZDC}$ data by a
sharp decrease of the $J/\psi$ multiplicity at small impact
parameter.

It is, however, not unlikely that the `second drop' in the
$E_{ZDC}$ data is an artifact of the data distortion. Indeed,
Fig.1 of Ref.\cite{NA50} shows clear disagreement of the
minimum bias distribution with the Glauber model at low $E_{ZDC}$.
This, as was explained above, invalidates the minimum bias
analysis procedure. Therefore, the apparent agreement between the
$E_T$- and $E_{ZDC}$ data in high centrality region obtained
within our analysis (see Figs.\ref{figE_T} and \ref{figE_ZDC}) may
be purely accidental.  In this case, there would be no
contradiction with the models
\cite{capellaD,Mai,Ko:02,Grandchamp:02}. Hence, refining of the
$E_{ZDC}$ data is very important. It would be hopefully
able to narrow the set of successful models in this field, which 
recently became too broad.

In conclusion, we performed a model independent comparison of two
data sets published by the NA50 collaboration: the $E_T$- and
$E_{ZDC}$-dependence of the $J/\psi$ to Drell-Yan ratio $R$ in
Pb+Pb collisions at SPS. We have shown that the two sets are at
variance. This problem seems to be related only to the data
obtained within the minimum bias procedure. A possible reason
for the problem may be a distortion of the minimum bias data
sample.  We argue that refining of the data is important to
clarify the qualitative and quantitative behavior of the ratio $R$
at $E_{ZDC} \lesssim 9$ TeV.

\ack We acknowledge the financial support of the Alexander von
Humboldt Foundation and GSI, Germany.

\begin{table}[p]
\begin{center}
\label{tab1}
\caption{The value of $\chi^2$ per experimental point at different
values of the number of nodes $J$ and for two different ways of choosing the
nodes (a) and (b) (see text for details).}
\vspace*{5mm}
\begin{tabular}{||r||c|c||c|c||c|c||}
\hline \hline
\multicolumn{1}{||c||}{J} &
\multicolumn{2}{c||}{$E_T$ fit} &
\multicolumn{2}{c||}{$E_{ZDC}$ fit} &
\multicolumn{2}{c||}{Combined fit} \\
\cline{2-7}
 & (a) & (b) & (a) & (b) & (a) & (b) \\
\hline \hline
  3 & 6.62 & 3.93 & 21.6 & 69.9 & 25.5 & 51.8 \\
  6 & 2.07 & 1.60 & 1.31 & 6.19 & 3.96 & 6.72 \\
 12 & 0.84 & 0.83 & 0.86 & 1.14 & 3.44 & 3.46 \\
 25 & 0.69 & 0.68 & 0.82 & 0.83 & 3.32 & 3.34 \\
 50 & 0.67 & 0.67 & 0.81 & 0.81 & 3.29 & 3.29 \\
100 & 0.67 & 0.67 & 0.80 & 0.80 & 3.28 & 3.29 \\
200 & 0.67 & 0.66 & 0.80 & 0.80 & 3.28 & 3.28 \\
\hline \hline
\end{tabular}
\end{center}
\vspace*{5mm}
\end{table}

\section*{References}

\begin{thebibliography}{99}

\bibitem{NA50}
M.~C.~Abreu {\it et al.}  [NA50 Collaboration],
Phys.\ Lett.\ B {\bf 521} (2001) 195.

\bibitem{comm}
A.~P.~Kostyuk, H.~Stocker and W.~Greiner,
arXiv:nucl-ex/0207018.

\bibitem{anomalous}
M.~C.~Abreu {\it et al.}  [NA50 Collaboration],
Phys.\ Lett.\ B {\bf 410} (1997) 337.

\bibitem{threshold}
M.~C.~Abreu {\it et al.}  [NA50 Collaboration],
Phys.\ Lett.\ B {\bf 450} (1999) 456.

\bibitem{evidence}
M.~C.~Abreu {\it et al.}  [NA50 Collaboration],
Phys.\ Lett.\ B {\bf 477} (2000) 28.

\bibitem{capella}
A.~Capella and D.~Sousa,
arXiv:nucl-th/0110072.

\bibitem{Chaurand}
B.~Chaurand, quoted in Ref. \cite{Mai}.

\bibitem{Ramello}
L.~Ramello   [NA50 Collaboration],
Talk given to the International Conference ``Quark Matter 2002'', Nantes,
France, July 18-24, 2002.

\bibitem{capellaD}
A.~Capella, E.~G.~Ferreiro and A.~B.~Kaidalov,
Phys.\ Rev.\ Lett.\  {\bf 85} (2000) 2080
[arXiv:hep-ph/0002300];\\
A.~Capella, A.~B.~Kaidalov and D.~Sousa,
Phys.\ Rev.\ C {\bf 65} (2002) 054908
[arXiv:nucl-th/0105021].

\bibitem{Mai}
J.~Blaizot, M.~Dinh and J.~Ollitrault,
Phys.\ Rev.\ Lett.\  {\bf 85} (2000) 4012 [nucl-th/0007020].

\bibitem{Ko:02}
A.~P.~Kostyuk, M.~I.~Gorenstein, H.~Stocker and W.~Greiner,
J.\ Phys.\ G {\bf 28} (2002) 2297
[arXiv:hep-ph/0204180];\\
A.~P.~Kostyuk,
arXiv:hep-ph/0209139.

\bibitem{Grandchamp:02}
L.~Grandchamp and R.~Rapp,
Nucl.\ Phys.\ A {\bf 709} (2002) 415
[arXiv:hep-ph/0205305];
arXiv:hep-ph/0209141.

\end{thebibliography}
\end{document}